\documentclass{article}
\usepackage{graphicx}
\usepackage{epsf}
\usepackage{epsfig}
\usepackage{amssymb}
\usepackage{amsmath}
\makeatletter
\@addtoreset{equation}{section}
\makeatother

\def\bea{\begin{eqnarray}}
\def\eea{\end{eqnarray}}
\def\be{\begin{equation}}
\def\ee{\end{equation}}

\title{Exact enumeration of Hamiltonian circuits, walks, and chains
       in two and three dimensions}
\author{Jesper Lykke Jacobsen${}^{1,2}$ \\[2.0mm]
${}^1$ LPTMS, Universit\'e Paris-Sud, B\^atiment 100, \\
Orsay, 91405, France \\
${}^2$ Service de Physique Th\'eorique, CEA Saclay, \\
Gif Sur Yvette, 91191, France}

\begin{document}

\maketitle

\begin{abstract}

We present an algorithm for enumerating exactly the number of Hamiltonian
chains on regular lattices in low dimensions. By definition,
these are sets of $k$ disjoint paths whose union visits each lattice
vertex exactly once. The well-known Hamiltonian circuits and walks appear
as the special cases $k=0$ and $k=1$ respectively. In two dimensions, 
we enumerate chains on $L \times L$ square lattices up to $L=12$,
walks up to $L=17$, and circuits up to $L=20$. Some results for three
dimensions are also given. Using our data we extract several quantities
of physical interest.

\end{abstract}

% \bigskip
% 
% \noindent SPhT-T07/XXX
%
% \smallskip

\section{Introduction}

The subject of Hamiltonian circuits and walks plays an important role
in mathematics and physics alike. Given a connected undirected graph
$G$, a {\em Hamiltonian circuit} (or cycle) is a cycle (i.e., a closed
loop) through $G$ that visits each of the $V$ vertices of $G$ exactly
once \cite{Hamilton}. In particular, a Hamiltonian circuit has length
$V$.  Similarly, a {\em Hamiltonian walk} (or path) is an open
non-empty path (i.e., with two distinct extremities) of length $V-1$
that visits each vertex exactly once. Note that a Hamiltonian circuit
can be turned into a Hamiltonian walk by removing any one of its
edges, whereas a Hamiltonian walk can be extended into a Hamiltonian
circuit only if its end points are adjacent in $G$.

\begin{figure}
  \centering
  \includegraphics[width=100pt,angle=0]{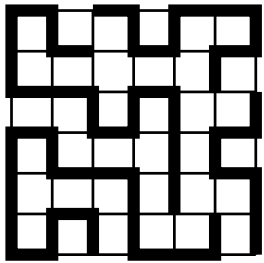}
  \caption{Hamiltonian chain of order 4 on a square lattice of size $7 \times
    7$.}
 \label{fig:chain}
\end{figure}

We add now to this list of well-known definitions the set ${\cal C}_k$ of {\em
  Hamiltonian chains} of order $k$. Each member in ${\cal C}_k$ is a set of
$k$ disjoint paths whose union visits each vertex of $G$ exactly once (see
Fig.~\ref{fig:chain}). The set of Hamiltonian walks is then ${\cal C}_1$, and
by convention we shall let ${\cal C}_0$ denote the set of Hamiltonian
circuits. Note that if $V$ is even, ${\cal C}_{V/2}$ is the set of dimer
coverings of $G$. The Hamiltonian chain problem has been studied earlier by
Duplantier and David on the Manhattan lattice \cite{DupDav}, but never to our
knowledge on an undirected lattice.

Determining whether $G$ contains a Hamiltonian circuit is a difficult
(NP-complete) problem. An even more difficult problem is to determine
{\em how many} distinct Hamiltonian circuits are contained in $G$.  In
this paper we shall present an algorithm that efficiently enumerates
Hamiltonian circuits, walks, and chains for regular low-dimensional
graphs.

The motivation for studying such Hamiltonian structures is by no means
limited to graph theory. Indeed, under appropriate solvent conditions,
biopolymers such as proteins may fold to form compact conformations,
the study of which is currently at the centre of an intense activity
in the biophysics community. While real biopolymers contain
complicated interactions which can probably not be fully accounted for
within any simple lattice model, the study of Hamiltonian walks has
been advocated as a first approximation for understanding
qualitatively the excluded-volume mechanisms at work behind such
problems as polymer melting \cite{Flory} and protein folding
\cite{Dill}. Our extension to Hamiltonian chains permits to study
polydisperse models of several polymers.

Another interest stems from the study of magnetic systems with O($n$) symmetry
in physics. These can modelled on the lattice as self-avoiding loops (each
having the weight $n$) \cite{Nienhuis}, which, in the limit of vanishing
temperature $T$, are constrained to visit all the vertices \cite{Kondev}.
Coupling such systems to a magnetic field $H$ amounts, in a perturbative
expansion around $H=0$, to inserting pairs of loop end points \cite{deGennes}.
In the limit $n\to 0$, the partition function $Z$ of the O($n$) model at $T=0$
in a weak magnetic field $H$ can thus be expressed in terms of the number of
Hamiltonian chains as
\begin{equation}
 Z = \sum_k {\cal C}_k H^{2k} \,.
\label{partfunc}
\end{equation}

Finally, the exact enumeration of configurations is useful for
settling issues of ergodicity when developing algorithms that provide
unbiased sampling of Hamiltonian walks in two \cite{Oberdorf} and
three \cite{Ham3D} dimensions.

In section~\ref{sec:alg} we present our enumeration algorithm and
discuss some aspects of its implementation. Results in dimensions
$d=2$ and $d=3$ are given in section~\ref{sec:res}. For convenience,
we limit the discussion to the simplest lattices (square and cubic),
although the construction extends straightforwardly to any regular
lattice. Our results are strongest in $d=2$ where we determine all
${\cal C}_k$ for $L \times L$ square lattices up to $L=12$, ${\cal
  C}_1$ up to $L=17$, and ${\cal C}_0$ up to $L=20$. In $d=3$ the
largest lattice that we were able to access has size $3 \times 4
\times 4$. Finally, we show in section~\ref{sec:app} how to extract
physically interesting quantities from our data.

\section{Algorithm}
\label{sec:alg}

We first present our algorithm in dimension $d=2$ and then discuss the
necessary modifications for $d=3$. For convenience, we limit the
presentation to the simplest lattices, viz.~an $L_1 \times L_2$ square
lattice and an $L_1 \times L_2 \times L_3$ cubic lattice, although
the construction extends straightforwardly to any regular lattice.
The boundary conditions are free (non-periodic), although it will be
clear that it is easy to introduce periodic boundary conditions along
{\em one} of the lattice directions.

The algorithm is based on the transfer matrix principle, according to
which the lattice is cut into two parts by means of a
conveniently chosen $d-1$ dimensional oriented surface ${\cal S}$. The
part of the lattice above (resp.\ below) ${\cal S}$ is called the
future (resp.\ the past). The surface ${\cal S}$ cuts the lattice only
at mid points of edges.

\begin{figure}
  \centering
  \includegraphics[width=400pt,angle=0]{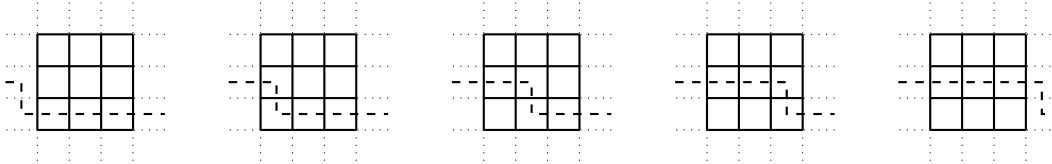}
  \caption{Transfer process for $d=2$. The lattice is shown in solid line
 style, with the dotted lines representing lattice edges beyond the bounds
 of the lattice. The surface ${\cal S}$ is shown as a dashed line.}
 \label{fig:transfer}
\end{figure}

At the initial (resp.\ final) step of the enumeration the whole
lattice belongs to the future (resp.\ past), so the algorithm consists
in sweeping ${\cal S}$ over the entire lattice.  This is done by
gradually pushing ${\cal S}$ towards the future, so that in any one
step of the algorithm a {\em single} vertex is transferred from the
future to the past. A few subsequent steps for a $4 \times 4$ lattice
are shown in Fig.~\ref{fig:transfer}.

\begin{figure}
  \centering
  \includegraphics[width=100pt,angle=0]{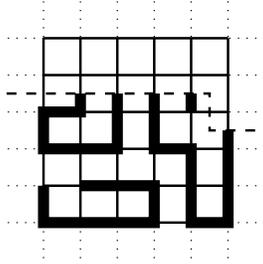}
  \caption{Configuration on a partially constructed $6 \times 6$ square.}
 \label{fig:config}
\end{figure}

In any step, the configuration of the system is described by some information
about the edges cut by ${\cal S}$, and by the number $k$ of chains which have
already been completed. The information refered to is the connectivity of the
cut edges with respect to the part of the lattice which belongs to the past,
and is best illustrated by an example (see Fig.~\ref{fig:config}). The
complete description of the configuration reads in this case
$(0,1,1,2,3,2|0|1)$, where the first $L_1$ entries refer to the state of the
cut edges which are parallel to the 2-direction (vertical), and the next entry
refers to the state of the one cut edge which is parallel to the 1-direction
(horizontal). The last entry is the number $k=1$ of completed chains. In the
connectivity part of the information, we use the following coding:
\begin{enumerate}
 \item A zero entry means an empty edge.
 \item Two equal positive entries mean a pair of edges which are connected in
   the past by part of a chain. Each of these edges will eventually have to be
   linked to a chain end point in the future, so as to form a complete chain.
 \item An unpaired positive entry means an edge which forms part of a
   partially completed chain, one end point of which has already been fixed in
   the past. The edge will eventually be linked to another end point in the
   future, leading to the formation of a complete chain.
\end{enumerate}
It is important to avoid any redundancy in this connectivity information.
A unique coding is obtained by requiring that the positive entries (i.e.,
unpaired entries, or the leftmost member of a pair of equal entries) be
arranged in increasing order ($1,2,3,\ldots$) when reading through the coding
from left to right.

In step $t$ of the enumeration, the configurations are transfered from
``time'' $t$ to time $t+1$. More precisely, each configuration at time $t$ is
examined in turn, and all its descendent configurations at time $t+1$ are
generated by exhausting the possible arrangements of the chain at the vertex
which is transfered from the future to the past. The information described
above is necessary and sufficient for deducing the connectivity information at
time $t+1$ from that at time $t$. Note that since the transfered vertex is not
allowed to be empty, there are four possible arrangements if it accommodates a
chain end, and six arrangements if a chain passes though it.

Each configuration generated at time $t+1$ is inserted in an
appropriate date structure---a hash table---which also keeps track of
its weight (here an integer). The weight of a descendent configuration
is the sum of the weights of all the parent configurations that
generated it. Some of the generated descendent configurations are
however {\em rejected} before insertion in the hash table (see below).
Once step $t$ has been completed, the hash table storing the
configurations at time $t$ is erased, and one can move on to step
$t+1$. In this way, only two hash tables (at times $t$ and $t+1$) are
needed in the entire process.

Note that the choice of data structure is essential for the feasibility
and the efficiency of the algorithm. A hash table permits to store the
configurations via a key which is obtained by reading its coding as one
large integer, modulo a suitably chosen prime. Storing and retrieving
configurations can be done in constant time, i.e., independently of the
number of configurations being stored in the hash table.

The hash table also allows to keep track of the weight of each configuration,
according to the above rule. Namely, when a descendent configuration is
generated with weight $w$, we first make an attempt of looking it up in the
hash table at time $t+1$. If it is not there, it is inserted with weight $w$.
If it is already there, $w$ is added to the weight of configuration already
present.

If in the transfer process the two ends of the same chain (coded by two
equal positive entries) join up, the resulting configuration is
rejected, since this would mean forming a cycle rather than a chain.
(We make an exception to this rule when the very last vertex is added,
since this permits to enumerate ${\cal C}_0$.) If an unpaired positive
entry gets left behind in the past it means that a chain has been
completed, and so $k \to k+1$.

Denote now a general configuration as
$(s_{2,1},s_{2,2},\ldots,s_{2,L_1}|s_1|k)$. At step $t=0$, the initial state
is $(0,0,\ldots,0|0|0)$ and has weight $1$. When a row of the lattice is
completed, any configuration with $s_1 \neq 0$ gets rejected. When transfering
the $i$'th vertex of the last row, any configuration with $s_{2,i} \neq 0$
gets rejected. This trick allows us to avoid having to deal with a lot of
special cases when a boundary vertex is transfered---and also makes it much
easier in practice to implement the algorithm correctly. After step $t=L_1 L_2$
the lattice belongs completely to the past, and all the configurations are of
the form $(0,0,\ldots,0|0|k)$. Their respective weights are precisely the
${\cal C}_k$ that we wanted to compute.

The maximum lattice size that we can attain is essentially limited by
the number of different intermediate configurations generated in the
transfer process. This number attains its largest value after
transfering the next last vertex in the next last row. In practice we
could store at most $\sim 10^8$ configurations.

As usual in enumeration studies, the coefficients ${\cal C}_k$ are
much larger than the integers which are usually represented by a
computer ($\le 2^{32}$ for an unsigned integer on a 32-bit machine). We
therefore repeat the enumeration several times, computing each time
the result modulo different coprime integers ($2^{32}$, $2^{31}-1$,
$2^{31}-3$,\ldots), and reconstruct the true result in the end by using
the Chinese remainder theorem. Note that the use of modular arithmetics
is possible because the weights of configurations are constructed only
by successive additions of positive integers.

The counts for systems of size $L_1 \times L_2$ and $L_2 \times L_1$ should of
course coincide. Verifying that this is indeed the case is however a very
strong check of the algorithm, since permuting $L_1$ and $L_2$ (with $L_1 \neq
L_2$) leads to a completely different transfer process in terms of the
propagation of the surface ${\cal S}$. We have performed such checks both
for $d=2$ and $d=3$.

We now describe briefly how the algorithm can be adapted to a
$d$-dimensional hypercubic lattice of size $L_1 \times L_2 \times
\cdots \times L_d$.  The surface ${\cal S}$ is pushed though the
lattice by means of $d$ nested loops, of which the innermost loop (at
nesting level $d-1$) moves ${\cal S}$ along the 1-direction, etc., and
the outermost loop (at nesting level $0$) moves ${\cal S}$ along the
$d$-direction. In general the loop at nesting level $\ell$ moves
${\cal S}$ along the $(d-\ell)$-direction.

A configuration is given by $(\{s_d\}|\{s_{d-1}\}|\cdots|\{s_1\}|k)$,
where the space $\{s_\ell\}$ describes the edges cut by ${\cal S}$
which are parallel to the $\ell$-direction and consists of
$\prod_{i=1}^{\ell-1} L_i$ entries. When the loop at nesting level
$d-\ell$ is executed for the last time, the entry in $\{s_\ell\}$
corresponding to the position of the loops at nesting levels $>d-\ell$
must be zero; otherwise the configuration is rejected.

At each vertex there are $2d$ possible local arrangements if the vertex
contains a chain end, and ${2d \choose 2}$ arrangements if it does not.

We have implemented the algorithm for $d=2$ and $d=3$. It is of course
most efficient in low dimensions when the number of entries necessary
to describe a configuration is small, and the configurations
themselves are strongly constrained by topology. We were however able
to obtain useful results as well for $d=3$ (see below).

\section{Results}
\label{sec:res}

\subsection{Two dimensions}

We first present our results in two dimensions.

\begin{table}
 \begin{center}
 \begin{tabular}{rr}
 $k$ & ${\cal C}_k$ \\ \hline
  0 & 1076226888605605706 \\
  1 & 3452664855804347354220 \\
  2 & 4105040900990127258563352 \\
  3 & 1716401559105599779087093260 \\
  4 & 363652056217217171206243035340 \\
  5 & 46148041957435926988244999692732 \\
  6 & 3870392655399966034741749180958852 \\
  7 & 229000444797839686805213214595470648 \\
  8 & 10014026193777241299766692880686035774 \\
  9 & 335126632781634776435981605808153310160 \\
 10 & 8818298873873444121262995871243309826506 \\
 11 & 186425336415902384343216389461927330172268 \\
 12 & 3222564357088784934867009058887596660853042 \\
 13 & 46216822292126998413476281985324396507245748 \\
 14 & 556693787783862608927984698386363470187981938 \\
 15 & 5690759125611797657956588062161464969526622268 \\
 16 & 49812855339352875263449851394183743992177894504 \\
 17 & 376259523557799790490076563092079957367971441020 \\
 18 & 2469068810121023544317004188530710352090945411914 \\
 19 & 14159258603854781892361610528528465090872582482860 \\
 20 & 71328606843660293099526723817890226149324996051770 \\
 21 & 317097933449802440642304292096463208253752002853220 \\
 22 & 1249084471204154623059161399602853755323161718519772 \\
 23 & 4375441585007318378166364769735565649135514407725428 \\
 24 & 13673227330703731913694049643952596405766113534103766 \\
 25 & 38227041578177782993926722638883248316303238609485072 \\
 26 & 95854543129195970883751499759209798433629811140560618 \\
 27 & 216054324315261909273142199235822975041539861478121508 \\
 28 & 438604247185012333557530043782600945755207568230643594 \\
 29 & 803320261178156739715403163647291125863198470203538048 \\
 30 & 1329413609161899137884978107235777594261127805779481138 \\
 31 & 1990427307505969944020236847124806666399811554129750496 \\
 32 & 2699118726122396972411170136925985393885382039591724116 \\
 33 & 3318040996252249925481740087801704155429556174343699608 \\
 34 & 3700359327341155923554786919768922871490843218081792948 \\
 35 & 3745875480760985174397430937750746156033853657449987776 \\
 \end{tabular}
 \end{center}
 \caption{Number ${\cal C}_k$ of Hamiltonian chains of order $k$ (i.e.,
   consisting of precisely $k$ chains) on a $12 \times 12$ square lattice.}
 \label{tab:all}
\end{table}

\begin{table}
 \addtocounter{table}{-1}
 \begin{center}
 \begin{tabular}{rr}
 $k$ & ${\cal C}_k$ \\ \hline
 36 & 3443370502558807290562354000836385732268208485521139030 \\
 37 & 2875008789367922659848083522243722363355224468273444220 \\
 38 & 2180493763475645200773617970678658343878870756101785154 \\
 39 & 1502095391370671575772987130076864642955725212507397512 \\
 40 & 939647724819716693388579329116456350135042650270872340 \\
 41 & 533562853839154855341927965097210146851707481332264436 \\
 42 & 274862642496699290252290990160279057175930352324091824 \\
 43 & 128362555399743719434877352579592630895860764794774624 \\
 44 & 54294968510865162278386828132399087185256262942414932 \\
 45 & 20778050812329434163278524326813392470106157931284728 \\
 46 & 7184786845784650425312709323260198674084909532776292 \\
 47 & 2241482341086337065894490074920307755810306693310936 \\
 48 & 629823815715953133331560488991904014678562091268042 \\
 49 & 159077245534404479393324552931624000438466735966020 \\
 50 & 36035273248074550220433523348607010222299409542106 \\
 51 & 7302543915028881257299361934945060798679283726808 \\
 52 & 1320074215106942124873659331692519827058055405644 \\
 53 & 212172409136637900106629576229565529242562898136 \\
 54 & 30209940418516563152074763182692186029320843454 \\
 55 & 3794668249506100568268876660836383536523653816 \\
 56 & 418512747239130986282686885840526291044063692 \\
 57 & 40310331424536568202368529058344353430213252 \\
 58 & 3369925505474814797946214848424889586336952 \\
 59 & 242794327043218279242923744053680409745316 \\
 60 & 14951909347540490900851120307089218715790 \\
 61 & 779493647407716363808229900734562711356 \\
 62 & 34012688096545799769214778699372499928 \\
 63 & 1225343373236044119139709333819722548 \\
 64 & 35847082988635315875798770248538788 \\
 65 & 834223151185015104742566527439696 \\
 66 & 15044012072281443196476714201368 \\
 67 & 203123129951305502546186136496 \\
 68 & 1958918610759335996516705296 \\
 69 & 12602334728369293472453184 \\
 70 & 48539905585658564517760 \\
 71 & 91904378228899701504 \\
 72 & 53060477521960000 \\
 \end{tabular}
 \end{center}
 \caption{(continued).}
\end{table}

We have been able to solve the full Hamiltonian chain problem for $L
\times L$ square lattices up to size $L=12$. The number of circuits
${\cal C}_0$ vanishes when $L$ is odd by an easy parity argument, but
the remaining ${\cal C}_k$ with $k=1,2,\ldots,\lfloor L^2/2 \rfloor$
are all non-zero. The complete result for $L=12$ is shown in
Table~\ref{tab:all}.

When $L$ is even, ${\cal C}_{L^2/2}$ should be the number ${\cal D}_L$
of dimer coverings of an $L \times L$ square lattice. We have checked
that our data agree with the analytical results \cite{Kasteleyn} for
${\cal D}_L$ for all $L=2,4,6,8,10,12$.

\begin{table}
 \begin{center}
 \begin{tabular}{rr}
 $L$ & ${\cal C}_1$ \\ \hline
  2 & 4 \\
  3 & 20 \\
  4 & 276 \\
  5 & 4324 \\
  6 & 229348 \\
  7 & 13535280 \\
  8 & 3023313284 \\
  9 & 745416341496 \\
 10 & 730044829512632 \\
 11 & 786671485270308848 \\
 12 & 3452664855804347354220 \\
 13 & 16652005717670534681315580 \\
 14 & 331809088406733654427925292528 \\
 15 & 7263611367960266490262600117251524 \\
 16 & 662634717384979793238814101377988786884 \\
 17 & 66428994739159469969440119579736807612665540 \\
 \end{tabular}
 \end{center}
 \caption{Number ${\cal C}_1$ of Hamiltonian walks on an $L \times L$ square
   lattice, up to size $L=17$.}
 \label{tab:walks}
\end{table}

If only the first few ${\cal C}_k$ are needed, the enumeration can be
taken to larger sizes by rejecting all states
$(s_{2,1},s_{2,2},\ldots,s_{2,L_1}|s_1|k)$ with $k > k_{\rm max}$.

In particular, we have obtained the number ${\cal C}_1$ of Hamiltonian
walks up to size $L=17$; see Table~\ref{tab:walks}.  This extends the
$L \le 7$ results by Mayer {\em et al} \cite{Mayer} by ten new terms.
Note also that Jaeckel {\em et al} \cite{Jaeckel} have proposed a
Monte Carlo method for estimating ${\cal C}_1$ for larger $L$. These
authors obtain $1.3582 \cdot 10^7$ for $L=7$---that is 0.3 \% above
the exact result---and $2.7791 \cdot 10^9$ for $L=8$---that is 29 \%
below the exact result.

Variant Hamiltonian walks, constrained to have their end points on
diametrically opposite corners of an $L \times L$ square with $L$ even, have
been studied in \cite{Jensen}. Since this is technically an easier problem
than our unconstrained walks, the enumerations could be taken to size $L=34$.

\begin{table}
 \begin{center}
 \begin{tabular}{rr}
 $L$ & ${\cal C}_0$ \\ \hline
  2 & 1 \\
  4 & 6 \\
  6 & 1072 \\
  8 & 4638576 \\
 10 & 467260456608 \\
 12 & 1076226888605605706 \\
 14 & 56126499620491437281263608 \\
 16 & 65882516522625836326159786165530572 \\
 18 & 1733926377888966183927790794055670829347983946 \\
 20 & 1020460427390768793543026965678152831571073052662428097106 \\
 \end{tabular}
 \end{center}
 \caption{Number ${\cal C}_1$ of Hamiltonian circuits on an $L \times L$ square
   lattice, up to size $L=20$.}
 \label{tab:circuits}
\end{table}

Finally, we have obtained the number ${\cal C}_0$ of Hamiltonian
circuits up to size $L=20$; see Table~\ref{tab:circuits}. This extends
the $L \le 16$ data \cite{Sloane} by two new terms.

\subsection{Three dimensions}

\begin{table}
 \begin{center}
 \begin{tabular}{rrrrrr}
 $k$ & $2 \times 2 \times 2$ & $2 \times 2 \times 3$
     & $2 \times 3 \times 3$ & $3 \times 3 \times 3$
     & $3 \times 3 \times 4$ \\ \hline
 0 &   6 &   22 &     324 &           0 &         3918744 \\
 1 &  72 &  584 &   16880 &     2480304 &       677849536 \\
 2 & 204 & 4204 &  270756 &   104844792 &     40656040968 \\
 3 & 108 & 7604 & 1281376 &  1246834176 &    816646740296 \\
 4 &   9 & 4541 & 2507084 &  6520250088 &   7803954743412 \\
 5 &     &  852 & 2281064 & 17852434656 &  41553510978656 \\
 6 &     &   32 &  985354 & 27873228036 & 134709106959932 \\
 7 &     &      &  190580 & 25864316448 & 280608712776492 \\
 8 &     &      &   13834 & 14445212196 & 388267754276278 \\
 9 &     &      &     229 &  4798350687 & 363680422635635 \\
10 &     &      &         &   911288760 & 232420898624633 \\
11 &     &      &         &    91325100 & 101132591631452 \\
12 &     &      &         &     4119048 &  29594655770318 \\
13 &     &      &         &       57048 &   5683316575620 \\
14 &     &      &         &             &    687432832414 \\
15 &     &      &         &             &     48991382300 \\
16 &     &      &         &             &      1837669320 \\
17 &     &      &         &             &        29304199 \\
18 &     &      &         &             &          117805 \\
 \end{tabular}
 \end{center}
 \caption{Number ${\cal C}_k$ of Hamiltonian chains of order $k$
   on various cubic lattices of size $L_1 \times L_2 \times L_3$.
   Blank entries are zero.}
 \label{tab:3Dchains}
\end{table}

Our results for the full Hamiltonian chain problem on small $L_1 \times L_2
\times L_3$ parallelepipeds are given in Table~\ref{tab:3Dchains}. In addition
we find for the $3 \times 4 \times 4$ system a number of ${\cal C}_0 =
3777388236$ circuits and ${\cal C}_1 = 1073054619800$ walks.

Note that the transfer matrix method is essentially limited by the area
of the smallest cross section of the parallelepiped. It would thus be
possible to extend the enumerations to some systems with, say, $L_1=L_2=3$
and $L_3 \ge 5$, but we have chosen to focus here on close-to-cubic shapes
which are the most challenging.

The number of walks ${\cal C}_1$ has been much studied in the area of protein
research \cite{Shakhnovich,Pande}, whereas the numbers ${\cal C}_k$ with $k
\neq 1$ have to our knowledge not been considered previously. Note that the
works \cite{Shakhnovich,Pande} were based on direct enumeration, meaning that
in contrast to our transfer matrix method each individual conformation was
actually generated. The limitation of direct enumeration is thus the number of
conformations being counted, since the CPU time requirement is (at best)
proportional to this. Accordingly, Ref.~\cite{Pande} uses massively parallel
supercomputer facilities to access the $3 \times 4 \times 4$ system. By
contrast, our transfer matrix approach is limited rather by the memory than
the CPU time. Unfortunately, this memory limitation put the $4 \times 4 \times
4$ system just a little outside our reach. On the other hand, the counts for
the $3 \times 4 \times 4$ system were made very fast, in just a few minutes.

In Refs.~\cite{Shakhnovich,Pande} the counts were produced modulo the symmetry
group of the lattice. For the $2 \times 2 \times 2$ and the $3 \times 3 \times
3$ systems, our results for ${\cal C}_1$ come out as exactly 24 times those of
\cite{Shakhnovich}. For the $3 \times 3 \times 4$ and $3 \times 4 \times 4$
systems, our results for ${\cal C}_1$ are precisely 8 times those of
\cite{Pande}. This means that for each of these systems, all Hamiltonian walks
are unrelated by lattice symmetries.

\section{Applications}
\label{sec:app}

The enumerations reported above conceal many quantities of physical relevance.
We discuss here some of them.

\subsection{Conformational exponents}

The radius of gyration $R$ of a polymer of length $l \gg 1$ is expected to
scale like
\begin{equation}
 R \sim l^\nu
\end{equation}
where $\nu$ is a standard critical exponent \cite{deGennes}. For Hamiltonian
walks on an $L^d$ hypercube in $d$ dimensions, we have obviously $R \sim L$
and $l \sim L^d$, and so $\nu = 1/d$. Non-trivial information is however
contained in the number of circuits and walks, here both supposed to have
one marked monomer attached to a fixed point:
\begin{equation}
 \tilde{\cal C}_1 \sim \mu^l l^{\gamma-1} \,, \qquad
 \tilde{\cal C}_0 \sim \mu^l l^{-\nu d} \,.
\end{equation}
Here $\mu$ is the so-called connective constant, and $\gamma$ is another
critical exponent \cite{deGennes}. In our setup, the circuits are unmarked
and the end points of the walks are free to be anywhere on the lattice,
and so
\begin{equation}
 {\cal C}_1 / {\cal C}_0 \sim l^{\gamma+1} \sim L^{(\gamma+1)d} \,.
\end{equation}
In addition to this leading behaviour there are subdominant corrections
due to surface effects.

In two dimensions, we can extract results for $\mu$ and $\gamma$ using the
data in Tables~\ref{tab:walks}--\ref{tab:circuits}. Due to parity effects,
this is best done by working in terms of the ratios
$\frac{\eta(L+2)}{\eta(L)}$, where $\eta(L)={\cal C}_0(L)$ for even $L$ or
${\cal C}_1(L)$ for any parity in the case of $\mu$, and $\eta(L)={\cal
  C}_1(L)/{\cal C}_0(L)$ for even $L$ in the case of $\gamma$. The naive
approximants are further extrapolated using standard finite-size scaling
techniques. This gives
\begin{eqnarray}
  \mu &=& 1.473 \pm 0.001 \,, \nonumber \\
  \gamma &=& 1.042 \pm 0.003 \,.
\end{eqnarray}

Our estimate for $\mu$ is in good agreement with (but less accurate than) the
currently best known estimate \cite{Kondev}
\begin{equation}
 \mu = 1.472801 \pm 0.00001 \,.
\end{equation}
Note that the latter uses exact predictions from field theory for
the leading finite-size corrections, a scheme that we have not adopted here.
The constrained Hamiltonian walks considered in \cite{Jensen} led to a
consistent value for $\mu$. The long-standing history of numerical and
analytical estimates for $\mu$ of Hamiltonian walks can be found in the
introductions of \cite{Jaeckel,Kondev}.

Our estimate for $\gamma$ is a nice confirmation of the exact field theoretical
result \cite{Kondev}
\begin{equation}
 \gamma = \frac{117}{112} = 1.04464 \cdots
\label{exact_gamma}
\end{equation}
Previous numerical results, as discussed in \cite{Kondev} and references
therein, were obtained in a cylindric geometry and assumed certain results
of conformal field theory. The present estimate therefore furnishes a more
direct verification of the exact result (\ref{exact_gamma}).

\subsection{Contact probabilities}

As already mentioned in the introduction, a Hamiltonian circuit
can be turned into a Hamiltonian walk by removing any one of its
edges, whereas a Hamiltonian walk can be extended into a Hamiltonian
circuit only if its endpoints are adjacent. This implies that the
probability that the two end points of a walk are adjacent is
\begin{equation}
 p_{\rm adj} = \frac{{\cal C}_0 L^d}{{\cal C}_1} \,.
\end{equation}

\begin{table}
 \begin{center}
 \begin{tabular}{rr}
 $L$ & $p_{\rm adj}$ \\ \hline
 2 & 1.00000000000000000000 \\
 4 & 0.34782608695652173913 \\
 6 & 0.16826830842213579364 \\
 8 & 0.09819321919798768694 \\
10 & 0.06400435120127304050 \\
12 & 0.04488610346836087660 \\
14 & 0.03315398616246303584 \\
16 & 0.02545282308230371747 \\
 \end{tabular}
 \end{center}
 \caption{Probability $p_{\rm adj}$ that the two end points of a Hamiltonian
   walk on an $L \times L$ square lattice are adjacent.}
 \label{tab:adjacent}
\end{table}

In two dimensions, the $p_{\rm adj}$ for $L \times L$ square can be computed
from Tables~\ref{tab:walks}--\ref{tab:circuits}. The resulting numerical
values are displayed in Table~\ref{tab:adjacent}. Note that $p_{\rm adj}$
vanishes for odd $L$ (as it does on any bipartite lattice having an odd number
of vertices).

The values in Table~\ref{tab:circuits} have been used in \cite{Oberdorf}
to test that a certain Monte Carlo algorithm for producing Hamiltonian
walks did indeed give unbiased results.

Physically, on may argue that $p_{\rm adj}$ is proportional to the probability
that the two ends of an open polymer (walk) join so as to form a ring polymer
(circuit).

It is tempting to try similarly to construct from the ratio of ${\cal C}_1$
and ${\cal C}_2$ the probability that the conformation of two chains is such
that one end point of each are adjacent on the lattice. Unfortunately, this
is not possible, since certain two-chains (more precisely those in which an
end point of one chain is adjacent to {\rm both} end points of the other
chain) can be obtained from more than one one-chain by removing an internal
edge in the latter.

\subsection{Lee-Yang zeroes}

The study of phase transitions through the location of partition function
zeroes in the complex magnetic field plane was initiated by Yang and Lee
\cite{YangLee}. In particular, these authors established that for the Ising
model these zeroes lie on the unit circle in terms of the variable $x={\rm
  e}^{2H}$, i.e., they correspond to purely imaginary values of the field $H$.
The zero closest to the positive real axis is denoted ${\rm e}^{i \theta_{\rm
    c}}$, where $\theta_{\rm c}$ is the so-called Lee-Yang edge. At the
critical temperature, $\theta_{\rm c} \to 0$ in the thermodynamic limit, and
its finite-size scaling permits to access a critical exponent.
Another possible approach is to study the density of zeroes $g(\theta)$.
Creswick and Kim \cite{Creswick} have shown that at the critical temperature
$g(\theta) \sim |\theta|^{1/\delta}$ for $\theta \ll 1$.

We have studied the zeroes of the partition function (\ref{partfunc}) in
terms of the variable $y=-H^2$ for $L \times L$ squares with $L \le 12$.
For odd $L$ there is one trivial zero at $y=0$ (since ${\cal C}_0=0$),
and for even $L$ the two zeroes closest to $y=0$ form a complex conjugate
pair with an imaginary part that tends to zero as $L \to \infty$. Disregarding
these ``exceptional'' zeroes, all the remaining zeroes (for $L$ of any parity)
are found to lie on the positive real axis in the complex $y$-plane,
corresponding to purely imaginary $H$ as in the Lee-Yang theorem.

We now define a finite-$L$ approximation to the density of zeroes in the point
$y_n$ as $g(y_n)=\frac{1}{y_{n+1}-y_n}$, where we have arranged the zeroes of
$Z(L \times L)$ in increasing order $y_1 < y_2 < \ldots < y_N$.

\begin{figure}
  \centering
  \includegraphics[width=300pt,angle=270]{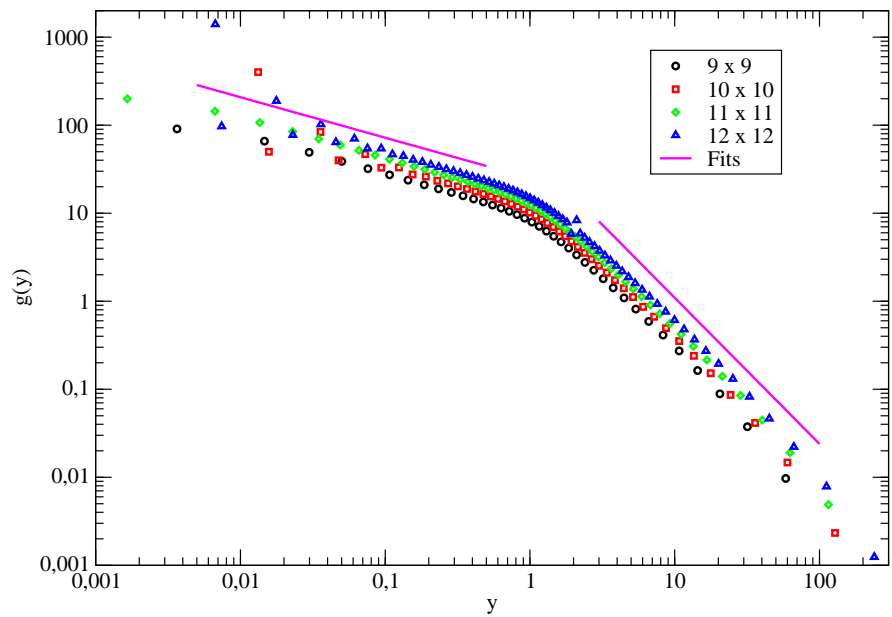}
  \caption{Density $g(y)$ of partition function zeroes in the variable $y=-H^2$
    for the Hamiltonian chain problem on $L \times L$ squares.}
 \label{fig:zeroes}
\end{figure}

The approximations $g(y)$ are shown in Fig.~\ref{fig:zeroes} for
$L=9,10,11,12$. One observes a clear crossover near $y=1$, separating two
regimes of power law behaviours. For even $L$ the curves bifurcate for $y \ll
1$, which can be remedied by regrouping the zeroes two by two (not shown). The
power laws extracted from the largest available sizes read
\begin{equation}
 g(y) \sim \left \lbrace \begin{array}{ll}
 y^{-0.46} & \mbox{for $y \ll 1$} \\
 y^{-1.66} & \mbox{for $y \gg 1$}
 \end{array} \right.
\end{equation}

We have no satisfying explanation for these exponents at present. The naive
application of the standard scaling laws $\nu d = 2 - \alpha$ (Josephson) and
$\alpha+2\beta+\gamma=2$ (Rushbrooke), and the results $\nu=\frac{1}{2}$ and
$\gamma=\frac{117}{112}$ for the critical $y=0$ system \cite{Kondev}, leads to
$1/\delta=-\frac{5}{229} = -0.0218\cdots$, which is clearly off the mark.

\section*{Acknowledgments}

This work was supported through the European Community Network ENRAGE (grant
MRTN-CT-2004-005616) and by the Agence Nationale de la Recherche (grant
ANR-06-BLAN-0124-03).


\begin{thebibliography}{99}

 \bibitem{Hamilton} W.R.~Hamilton, Phil.~Mag.~{\bf 12} (1856);
   Proc.~Roy.~Irish Acad.~{\bf 6} (1858).

 \bibitem{DupDav} B. Duplantier and F. David,
   J. Stat. Phys. {\bf 51}, 327 (1988).

 \bibitem{Flory} P.J. Flory,
   Proc. Roy. Soc. London A {\bf 234}, 60 (1956).

 \bibitem{Dill} K.A. Dill,
   Protein Science {\bf 8}, 1166 (1999).

 \bibitem{Nienhuis} B. Nienhuis,
   Phys. Rev. Lett. {\bf 49}, 1062 (1982).

 \bibitem{Kondev} J.L. Jacobsen and J. Kondev,
   Nucl. Phys. B {\bf 532}, 635 (1998).

 \bibitem{deGennes} P.-G. de Gennes, {\em Scaling concepts in polymer physics}
   (Cornell University Press, New York, 1979).

 \bibitem{Oberdorf} R. Oberdorf, A. Ferguson, J.L. Jacobsen and J. Kondev,
   Phys. Rev. E {\bf 74}, 051801 (2006).

 \bibitem{Ham3D} J.L. Jacobsen, preprint (2007).

 \bibitem{Kasteleyn} P.W. Kasteleyn, Physica {\bf 27}, 1209 (1961).

 \bibitem{Mayer} J.-M. Mayer, C. Guez and J. Dayantis,
    Phys. Rev. B {\bf 42}, 660 (1990).

 \bibitem{Jaeckel} A. Jaeckel, J. Sturm and J. Dayantis,
    J. Phys. A {\bf 30}, 2345 (1997).

 \bibitem{Jensen} M. Bousquet-M\'elou, A.J. Guttmann and I. Jensen,
    J. Phys. A {\bf 38}, 9159 (2005).

 \bibitem{Sloane} Sequence A003763 in N.J.A. Sloane (ed.), {\em The online
      encyclopedia of integer sequences}, {\tt
      http://www.research.att.com/$\widetilde{\ }$njas/sequences/}.

 \bibitem{Shakhnovich} E. Shakhnovich and A. Gutin,
   J. Chem. Phys. {\bf 93}, 5967 (1990).

 \bibitem{Pande} V.S. Pande, A.Y. Grosberg, C. Joerg and T. Tanaka,
   J. Phys. A {\bf 27} 6231 (1994).

 \bibitem{YangLee} C.N. Yang and T.D. Lee,
   Phys. Rev. {\bf 87}, 404 (1952); {\em ibid.} {\bf 87}, 401 (1952).

 \bibitem{Creswick} R.J. Creswick and S.-Y. Kim,
   Phys. Rev. E {\bf 56}, 2418 (1997).

\end{thebibliography}
\end{document}